\documentclass[unnumsec,webpdf,contemporary,large, table]{oup-authoring-template}%
\theoremstyle{thmstyleone}%
\theoremstyle{thmstyletwo}%
\theoremstyle{thmstylethree}%

\usepackage[table]{xcolor}
\usepackage{multirow}
\usepackage{subcaption}
\usepackage{array}

\usepackage{graphicx, booktabs}  % for inserting images
\graphicspath{{Fig/}}

\usepackage{tikz}
\usepackage{pgfplots}
\pgfplotsset{compat=1.18}
\usepgfplotslibrary{groupplots}
\usepackage{subcaption}
\pgfkeys{/pgf/number format/.cd,1000 sep={\,}}
\usepgfplotslibrary{fillbetween}

\usepackage{xcolor}
\usepackage{subfiles}

\definecolor{jublue1}{rgb}{0.204,0.392,0.69}

\usepackage{amsmath, amsfonts}
\usepackage{stmaryrd}  % for integer intervals
\usepackage{mathtools}  % for aligning signs in an equation

\newcommand{\mat}[1]{\mathbf{\MakeUppercase{#1}}}
\newcommand{\vect}[1]{\mathbf{\MakeLowercase{#1}}}
\newcommand{\norm}[1]{\left\lVert#1\right\rVert}
\newcommand{\A}{\mat{A}}
\newcommand{\R}{\mathbb{R}}
\newcommand{\G}{\mathcal{G}}

\newcommand{\minus}{\scalebox{0.75}[1.0]{$-$}}

\newcommand{\beginsupplement}{%
\setcounter{table}{0}
\renewcommand{\thetable}{S\arabic{table}}%
\setcounter{figure}{0}
\renewcommand{\thefigure}{S\arabic{figure}}%
}
     
\begin{document}

\journaltitle{Journal Title Here}
\DOI{DOI HERE}
\copyrightyear{2024}
\pubyear{2024}
\access{Advance Access Publication Date: Day Month Year}
\appnotes{Paper}

\firstpage{1}

%\subtitle{Subject Section}

\title[Gene Expression Profiling By Statistical and Machine Learning]{A Comparative Analysis of Gene Expression Profiling by Statistical and Machine Learning Approaches}

\author[1, 2]{Myriam Bontonou\ORCID{0000-0002-0010-5457}}
\author[3]{Anaïs Haget}
\author[3]{Maria Boulougouri}
\author[2, $\ast$]{Benjamin Audit\ORCID{0000-0003-2683-9990}}
\author[2, $\ast$]{Pierre Borgnat\ORCID{0000-0003-4536-8354}}
\author[1, $\ast$]{Jean-Michel Arbona\ORCID{0000-0001-6166-9056}}

\authormark{Bontonou et al.}

\address[1]{\orgdiv{Laboratoire de Biologie et Modélisation de la Cellule}, \orgname{ENS de Lyon, CNRS, UMR5239, Inserm U1293, Univ Lyon}, \orgaddress{\country{France}}}
\address[2]{\orgname{Univ Lyon, ENS de Lyon, CNRS}, \orgdiv{Laboratoire de Physique}, \orgaddress{F-69342 Lyon, \country{France}}}
\address[3]{\orgdiv{LTS2 laboratory}, \orgname{EPFL}, \orgaddress{\state{Lausanne}, \country{Switzerland}}}

\corresp[$\ast$]{Corresponding authors. \href{email:jean-michel.arbona@ens-lyon.fr}{jean-michel.arbona@ens-lyon.fr}, \href{email:benjamin.audit@ens-lyon.fr}{benjamin.audit@ens-lyon.fr}, \href{email:pierre.borgnat@ens-lyon.fr}{pierre.borgnat@ens-lyon.fr}}

\received{Date}{0}{Year}
\revised{Date}{0}{Year}
\accepted{Date}{0}{Year}

%\editor{Associate Editor: Name}

%\abstract{
%\textbf{Motivation:} .\\
%\textbf{Results:} .\\
%\textbf{Availability:} .\\
%\textbf{Contact:} \href{name@email.com}{name@email.com}\\
%\textbf{Supplementary information:} Supplementary data are available at \textit{Journal Name}
%online.}

\abstract{
Many machine learning models have been proposed to classify phenotypes from gene expression data. In addition to their good performance, these models can potentially provide some understanding of phenotypes by extracting explanations for their decisions. 
These explanations often take the form of a list of genes ranked in order of importance for the predictions, the highest-ranked genes being interpreted as linked to the phenotype.
We discuss the biological and the methodological limitations of such explanations. Experiments are performed on several datasets gathering cancer and healthy tissue samples from the TCGA, GTEx and TARGET databases. A collection of machine learning models including logistic regression, multilayer perceptron, and graph neural network are trained to classify samples according to their cancer type. 
Gene rankings are obtained from explainability methods adapted to these models, and compared to the ones from classical statistical feature selection methods such as mutual information, DESeq2, and EdgeR. 
Interestingly, on simple tasks, we observe that the information learned by black-box neural networks is related to the notion of differential expression.
In all cases, a small set containing the best-ranked genes is sufficient to achieve a good classification. However, these genes differ significantly between the methods and similar classification performance can be achieved with numerous lower ranked genes.
In conclusion, although these methods enable the identification of biomarkers characteristic of certain pathologies, our results question the completeness of the selected gene sets and thus of explainability by the identification of the underlying biological processes.
}

\maketitle

\section{Introduction}
A phenotype results from a complex cascade of molecular processes possibly involving hundreds or thousands of genes. Understanding the connections between phenotype and gene expression is crucial to better identify mechanisms implicated in diseases such as cancer. 

Statistical methods like EdgeR~\cite{robinson2010edger} and DESeq2~\cite{love2014moderated}, have been pivotal in analysing gene expression levels. They identify genes that are expressed differentially between phenotypes and suggest to select them as markers of these phenotypes. However, recent Machine Learning (ML) methods, having proved their usefulness in other fields \cite{Goodfellow-et-al-2016}, open a promising avenue to enhance diagnostic accuracy and unravel complex biological processes. This has been showcased for identifying cancer subtypes~\cite{golub1999molecular, hoadley2018cell, parker2009supervised} and for training diagnostic tools on large datasets~\cite{leng2022benchmark, hanczar2022assessment}, leading to the identification of discriminating sets of genes~\cite{li2017comprehensive, jacquet2023aberrant}. Various models, such as $k$-nearest neighbours~\cite{li2017comprehensive}, support vector machines~\cite{rohimat2022implementation}, deep neural networks~\cite{ahn2018deep} and graph neural networks~\cite{ramirez2020classification}, have been successfully trained to classify samples from their gene expression profiles. Discriminating gene sets can be determined either before, during, or after training~\cite{mahendran2020machine, alharbi2023machine}. 
Integrating these ML methods in the analysis of gene expression levels suggests a paradigm shift in these analyses: interpretability (or explainability) techniques for ML \cite{molnar2022} offer ways to explain the decisions of the models, and provide scores that allow to rank and possibly select relevant genes to explain specific phenotypes. However, there is a significant gap between the selection of discriminating gene sets and our understanding of the biological processes involved in the development of certain phenotypes, because distinct gene sets can yield similar classification performance~\cite{li2017comprehensive}.

Here, we study whether ML models and associated scores of explainability can unveil novel, biologically relevant, molecular signatures. The signatures obtained from several statistical and ML methods are studied in terms of their classification performance and their biological relevance. Biological relevance is evaluated through over-representation analysis (ORA)~\cite{khatri2012ten}, which computes the overlap of the top-ranked genes with established gene sets representing various biological processes. In this context, gene expression profiling refers to the creation of a profile of genes ranked according to their relevance for understanding (and possibly predicting) phenotypes.
To elucidate the specific insights that ML can offer, the top ranked genes selected by ML methods are compared with those selected by EdgeR~\cite{robinson2010edger} and DESeq2~\cite{love2014moderated} methods for differential gene expression.
Several classifiers, including logistic regressions, shallow neural networks leveraging more complex relationships, and graph-based neural networks exploiting gene interactions, are trained to differentiate cancer tissue samples. Genes rankings are derived from the ML classifiers using the integrated gradients explainability method (IG)~\cite{sundararajan2017axiomatic}.

Experiments are carried out on existing gene expression datasets from The Cancer Genome Atlas (TCGA)\footnote{\footnotesize{www.cancer.gov/tcga}},  the Therapeutically Applicable Research to Generate Effective Treatments programme (TARGET)\footnote{\footnotesize{www.cancer.gov/ccg/research/genome-sequencing/target}} and the Genotype-Tissue Expression project (GTEx)~\cite{lonsdale2013genotype}. 

The results, in line with existing literature, identify small gene sets maintaining classification performance. Still, the question of the relevance of genes selected through ML models explainability remains unsolved. We observe that the top-ranked genes identified by ML methods vary, and differ significantly from those identified by statistical methods --  presenting themselves significant variations. Interestingly, a classifier trained on genes selected by DESeq2 or mutual information outperforms one trained on genes specifically selected for it. Over-representation analysis reveals diverse biological processes, sometimes specific to a single method or a paired methods, suggesting different facets of explainability. In simple cases, the ranks obtained by explaining ML models predictions correlate well with t-statistic ranks comparing gene means of two classes, which is aligned with our intuitive understanding of the problem.

In summary, we undertake a comprehensive analysis of gene expression profiling using emerging ML techniques by questioning classification efficiency and biological relevance. Comparisons with traditional statistical methods provide insights into the evolving landscape of molecular signature identification. The code and data are accessible at {\small{\url{https://github.com/mbonto/XAI_in_genomics}}}.

\section{Materials and methods}
\subsection{Setting}
Let's consider a gene expression dataset containing $N$ training data samples of class $c\in\llbracket 1; C\rrbracket$. The class reflects the phenotypic state of a tissue. Here, it is a cancer type, a cancer subtype, or a healthy type. A data sample is a feature vector $\vect{x}\in\R^G$ containing the average expression of $G$ genes within a tissue sample. The true class $c$ of a sample is represented by a one-hot vector $\hat{\vect{y}}\in\R^C$ having value 0 everywhere except for $\hat{\vect{y}}_c = 1$. The Euclidean norm is written $\norm{\cdot}$.

\subsection{Machine learning models for classification}
To solve a classification task, a supervised model $f: \R^G \mapsto \R^C$ learns to map the features $\vect{x}$ of a data sample to a vector $\vect{y}\in\R^C$ whose coefficients $\vect{y}_c$ represent the probability of belonging to class $c$.
Three types of models are used in the present work: logistic regression (LR), multilayer perceptron (MLP), and graph neural network (GNN).

\subsubsection{Logistic regression (LR)}
LR models with L2 or L1 regularisation for both binary ($C=2$ classes) and multi-class ($C > 2$) classification problems are considered. 
In the binary case, a sigmoid function is applied to compute class probabilities~\cite{bishop2006pattern}. Given the parameters $\mat{W}\in\R^{1\times G}$, $b\in\R$ and $\sigma(z) =  1/(1 + \exp(\minus z))$ the sigmoid function, the model is $\vect{y}=\sigma(\mat{W}\vect{x} + b)$.
In the multi-class case, a softmax function is used instead of $\sigma$ for the probabilities~\cite{bishop2006pattern}.
Given the parameters $\mat{W}\in\R^{C\times G}$ and $\vect{b}\in\R^C$,
$
    \vect{y}=\text{softmax}(\mat{W}\vect{x} + \vect{b})\text{ with softmax}(\vect{z}_g) = \frac{\exp(\vect{z}_g)}{\sum_f\exp(\vect{z}_f)}\,.
$
Two versions of LR, LR+L2 and LR+L1, are considered. LR+L2 is trained with a L2 penalty, a regularisation term equal to the squared values of the parameters.
LR+L1 uses a L1 penalty term equal to the absolute values of the parameters; it has the advantage of sparsity. As the number of non-zero parameters is minimised, the interpretability of potential candidate genes relevant for phenotypes is enhanced.
These regularisation techniques both prevent the model from over-fitting to the training data, and improve the generalisation performance~\cite{bishop2006pattern}.

\subsubsection{Multilayer perceptron (MLP)}
A MLP is a neural network containing a series of fully connected layers followed by a LR~\cite{Goodfellow-et-al-2016}. 
Formally, $G^{[l]}$ is the number of hidden features after layer $l$ with $G^{[0]}=G$.
In the binary case, 
$
    f(\vect{x}) = \sigma(\vect{b}^{[L]} + \mat{W}^{[L]} \text{\small{ReLU}}(\dots \text{\small{ReLU}}(\vect{b}^{[1]} + \mat{W}^{[1]} \vect{x})))
$.
The parameters learned within layer $l$ are $\mat{W}^{[l]}\in\R^{G^{[l]}\times G^{[l-1]}}$ and $\vect{b}^{[l]}\in\R^{G^{[l]}}$. ReLU is the Rectified Linear Unit function.
For the multi-class case, the sigmoid of the last layer is replaced by softmax.
Here, the MLP architecture also includes a batch normalisation function to each layer, which stabilises training and improves generalisation~\cite{Goodfellow-et-al-2016}. In the experiments, shallow MLP with 1 or 2 layers are used (see hyperparameter selection).

\subsubsection{Graph neural network (GNN)}
A GNN~\cite{zhou2020graph} incorporates  gene pairs relationships using a graph structure. Following~\cite{ramirez2020classification}, a graph is created by connecting co-expressed genes through Pearson correlations.
This generates a graph $\G$, with nodes $\mathcal{V} = \left\{1, \dots, G \right\}$ representing genes and edges $\mathcal{E}$ denoting relationships between gene pairs. The edge weights (i.e., thresholded correlations) are stored in the adjacency matrix $\A \in \R^{G \times G}$ of $\G$. The applied threshold is discussed in the hyperparameter selection subsection.
As in~\cite{ramirez2020classification}, the GNN architecture includes 1 or 2 neural network layers followed by a LR. Each neural network layer comprises a graph convolutional layer~\cite{DBLP:conf/iclr/KipfW17} followed by a graph coarsening layer based on the Graclus algorithm~\cite{dhillon2007weighted}. 

Formally, $F[l]$ represents the number of features associated with a gene after layer $l$. In the initial layer, the gene expression value is the only feature ($F[0]=1$).
The first graph convolutional layer is $
\text{ReLU}(\vect{b}^{[1]} + \tilde{\mat{D}}^{-\frac{1}{2}}\tilde{\A}\tilde{\mat{D}}^{-\frac{1}{2}}\vect{x}\mat{W}^{[1]})$ with the data sample $\vect{x}\in \R^{G\times 1}$, the normalised adjacency matrix $\tilde{\A} = \A + \mat{I}$ and $\tilde{\mat{D}}$ the degree matrix of $\tilde{\A}$~\cite{DBLP:conf/iclr/KipfW17}. The same parameters $\mat{W}^{[1]}\in\R^{1\times F[1]}$ and $\vect{b}\in \R^{F[1]}$ are applied to all genes. The graph coarsening layer~\cite{dhillon2007weighted} combines pairs of adjacent nodes into a single node, preserving maximal features (max pooling). New edges result from the union of previous edges, with associated weights summed.

\subsection{Gene selection and explainability for gene profiling}
The objective is to identify biologically significant molecular signatures using gene expression data. Several methods assign a score $\boldsymbol{\phi}_g$ to each gene $g$ (its expression level being an input feature) indicating its relative importance for the sample phenotype class. 
These methods are categorised into three groups.
%\begin{enumerate*}[label=(\roman*)]
(i) \textit{Filter methods} rank genes based on a statistics measuring the amount of information they contain; filtering is done independently of any specific classifier.
(ii) \textit{Embedded methods} rank genes based on a score derived from a classifier; gene ranking is integrated into the training process of the classifier.
(iii) \textit{Post-hoc methods} rank genes based on a score computed after the training of a classifier; gene ranking occurs as a separate step after the classifier is trained.
%\end{enumerate*}

\subsubsection{Filter method - Variance (VAR)}
Genes can be ranked in decreasing order of their variance across samples in the dataset; filtering out genes with low variance is often standard practice when preprocessing a dataset. By focusing on the gene expression data only and ignoring their corresponding phenotypes, this ranking method disregards true classes. Let $\overline{\vect{x}}\in\R^{G}$ be the average on all the training samples. Then, the score is: 
\begin{equation}
    \boldsymbol{\phi}_g^\text{VAR} = \frac{1}{N}\sum_{n=1}^N (\vect{x}_{g}^n - \overline{\vect{x}}_g)^2\,.
\end{equation}

\subsubsection{Filter method - Principal component analysis (PCA)}
Genes can be ranked according to their contribution to the main directions of data variability. The higher the magnitude of a gene's coefficient in the principal components, the more significant it is~\cite{jolliffe2002principal}. PCA focuses on the variability of the dataset without considering the classes. As large proportion of this variability is caught by the first principal component $\vect{v}^1$, i.e. associated with the highest eigenvalue of the covariance $\mat{\Sigma}\in\R^{G \times G}$, only this one is kept here:: 
\begin{equation}
    \boldsymbol{\phi}_g^\text{PCA} = \lvert\vect{v}^1_g\rvert\,.
\end{equation}

\subsubsection{Filter method - Mutual information (MI)}
To account for higher order terms in probabilities, genes are ranked in decreasing order of the mutual information $I(X_g;Y)$ shared between the expression $X_g$ of a gene $g$,  and the classes represented by a discrete variable $Y$~\cite{ross2014mutual}. 
Then:
\begin{equation}
    \boldsymbol{\phi}_g^\text{MI} = I(X_g;Y)\,.
\end{equation}

\subsubsection{Filter method - Differential expression (EdgeR, DESeq2)}
Popular bioinformatics tools are used to identify differentially expressed genes between experimental conditions~\cite{costa2017rna}. Once the expression distribution of a gene is modelled, a test for differential expression is performed and yields a p-value adjusted to account for multiple testing.

In this study, we consider EdgeR~\cite{robinson2010edger} and DESeq2~\cite{love2014moderated}, two methods of this category that rely on different testing procedures. Genes are then ranked according to their adjusted p-values; low adjusted p-values indicating a high statistical significance for differential expression. When more than two classes are considered, a test is performed for each possible pair of classes. The genes are ranked based on the minimal p-value obtained across all these tests. This ranking strategy reflects the presence of a significant difference for at least one pair of classes. 
The scores are:
\begin{equation}
    \boldsymbol{\phi}_g^\text{EdgeR, DESeq2} = -\text{log}_{10}(\text{adjusted p-value}^\text{EdgeR, DESeq2}_g)\,.
\end{equation}

\subsubsection{Embedded method - Magnitude of the LR weights}
For LR, the magnitude of the weight $\mat{W}_g$ associated to a gene $g$ reflects its impact on the classification, as a larger one signifies a more substantial influence of the associated gene. If it has positive value, it indicates that over-expression favours the class, while negative value suggests that under-expression does so. For binary classification, an intrinsic score is directly the magnitude of this weight~\cite{molnar2022}.
As gene expressions are standardised, this metric ensures comparability of magnitudes across genes. The score for binary LR is:
\begin{equation}
    \boldsymbol{\phi}_g^\text{LR (weight)} = |\mat{W}_g|\,. 
\end{equation}

For multi-class, several linear functions (one for each class) are simultaneously learned. By analogy, genes are ranked based on the average of the absolute values of the parameters associated with a gene, hence:
\begin{equation}
    \boldsymbol{\phi}_g^\text{LR (weight)} = \frac{1}{C}\sum_{c=1}^C \lvert\mat{W}_{cg}\rvert\,. 
\end{equation}

\subsubsection{Post-hoc method - Integrated Gradients (IG)}
\label{sec:evaluation}
Neural networks are not directly interpretable because of their functional complexity~\cite{molnar2022}. To highlight the individual features that impact the most the decision of a neural network on a particular example, several explainability methods have been developed~\cite{lundberg2017unified}. Here, the integrated gradients method (IG)~\cite{sundararajan2017axiomatic} is chosen to explain the decisions made by various gradient-based ML models, specifically LR, MLP, and GNN. IG is a gradient-based technique that is used a lot in the literature, in particular for its computational efficiency.

IG assigns a score function $\boldsymbol{\phi}_g^{\text{local}}(\vect{x})$ to each gene $g$ in each sample $\vect{x}$ to represent its importance for the model's decision. The computation of the scores $\boldsymbol{\phi}_g^{\text{local}}(\vect{x})$ contrasts the prediction for a sample $\vect{x}$ with a reference sample $\vect{x}^\prime$ (baseline) according to:
\begin{equation*}
    \boldsymbol{\phi}_g^{\text{local}}(\vect{x}) = (\vect{x}_g - \vect{x}_g^\prime) \int_{\alpha=0}^1 \frac{\partial f_c(\vect{z})}{\partial \vect{x}_g} \bigg|_{\vect{z}=\vect{x}^\prime + \alpha(\vect{x} - \vect{x}^\prime)}d\alpha\,.
\label{eq:IGdef}
\end{equation*}
Here, $f_c(\vect{z})$ is the output of the model for class $c$ at input $\vect{z}$.
To aggregate the decisions of a model, a global score $\boldsymbol{\phi}_g^{\text{class}}$ for the importance of a gene $g$ within a class $c$ is derived using training samples that are correctly classified. If $N_\text{class}$ is the number of data samples of class $c$ used to train a model, then:
\begin{equation}
    \boldsymbol{\phi}_g^{\text{class}} = \frac{1}{N_\text{class}} \sum_{n=1}^{N_\text{class}}\frac{\lvert\boldsymbol{\phi}_g^{\text{local}}(\vect{x}^n)\rvert}{\norm{[\boldsymbol{\phi}_1^{\text{local}}(\vect{x}^n), \dots, \boldsymbol{\phi}_G^{\text{local}}(\vect{x}^n)]}}
\end{equation}
The normalisation ensures comparability across different samples. When multiple classes are studied, a global metric is derived by averaging the importance scores:
\begin{equation}
    \boldsymbol{\phi}_g^\text{IG} = \frac{1}{C}\sum_c\boldsymbol{\phi}_g^{\text{class}}\,. 
\end{equation}
The studied classes and the baselines used for each classification task considered in this article are detailed in the feature selection methods subsection of the Materials Section.

\subsubsection{Prediction Gaps for IG}
\begin{figure}[t]
\centering
\begin{tikzpicture}
\begin{groupplot}[group style={group size=1 by 2, horizontal sep=0cm, vertical sep=0cm}]

    \nextgroupplot[
    hide axis,
    width=5cm,
    height=2.5cm,
    ymin=0,
    ymax=10,
    xmin=0,
    xmax=10,
    legend style={font=\footnotesize, at={(1.3,0.8)},anchor=east, fill opacity=0.5, text opacity=1},
    column sep=0.25cm
    ]
    
    \addlegendimage{color=orange, line width=1pt}
    \addlegendentry{\small{Most important variables masked first}}; 
    \addlegendimage{color=jublue1, line width=1pt}
    \addlegendentry{\small{Less important variables masked first}};

    \nextgroupplot[
    width=8cm, 
    height=3cm,
    axis lines = left,
    enlargelimits = true, 
    xlabel=Proportion of masked variables (\%), 
    ymin=1, 
    ymax=110,
    xmin=0,
    xmax=100,
    xtick={0,10,...,100},
    ytick={0,50,...,100},
    ylabel near ticks,
    ylabel style={align=center},
    ylabel={Prediction error\\(\%)},]

    \addplot[name path=PGI, color=orange, mark=none, rounded corners=0pt, line width=1pt] table [x=Prop, y=PGI, col sep=comma] {csv/curves.csv};
    \addplot[name path=PGU, color=jublue1, mark=none, rounded corners=0pt, line width=1pt] table [x=Prop, y=PGU, col sep=comma] {csv/curves.csv};
    \addplot[color=gray, dashed, line width=2pt, mark=none, rounded corners=0pt] table [x=Prop, y=Default, col sep=comma] {csv/curves.csv};

    \path[name path=axis] (axis cs:0,0) -- (axis cs:100,0);
    \addplot [
        thick,
        color=orange,
        fill=orange, 
        fill opacity=0.05
    ]
    fill between[
        of=PGI and axis,
        soft clip={domain=0:100},
    ];

    \path[name path=axis] (axis cs:0,0) -- (axis cs:100,0);
    
    \addplot [
        thick,
        color=jublue1,
        fill=jublue1, 
        fill opacity=0.05
    ]
    fill between[
        of=PGU and axis,
        soft clip={domain=0:100},
    ];

    \node[jublue1, overlay] at (axis cs: 85, 43) {PGU};
    \node[orange, overlay] at (axis cs: 45, 43) {PGI};
    \node[black, overlay] at (axis cs: 7, 88) {\footnotesize $\frac{f_c(\vect{x}) - f_c(\vect{x}^\prime)}{f_c(\vect{x})}$};

\end{groupplot}
\end{tikzpicture}
\vspace{-0.8cm}
\caption{Illustration of PGI and PGU: prediction gaps on important (PGI) and unimportant (PGU) features, for an example $\vect{x}$ in a class $c$.}
\label{fig:PG_scheme}
\end{figure}
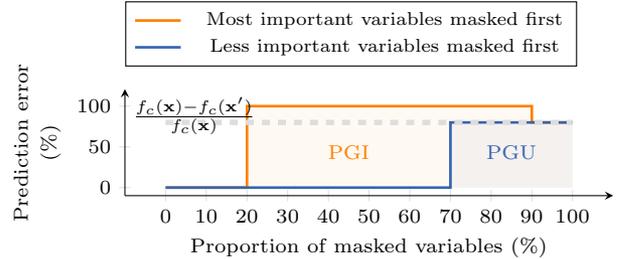

One can obtain more insights into the significance of the IG ranking in the classification process by studying the impact of progressively masking genes on the model's predictions.
This masking, referred to as \textit{experiment~0 in the following} is performed by replacing the gene expression values in the original sample $\vect{x}$ with the corresponding values from a reference sample $\vect{x}^\prime$. 
For a sample $\tilde{\vect{x}}_m$ with $m$ masked variables, the prediction gap (PG)~\cite{agarwal2022openxai} is:
\begin{equation*}
    \text{PG} = \frac{1}{G}\sum_{m=1}^{G} \frac{\max (f_c(\vect{x}) - f_c(\tilde{\vect{x}}_m), 0)}{f_c(\vect{x})}\;.
\end{equation*}
Genes can be masked using different orders on the predefined rankings $\boldsymbol{\phi}$: in descending order of importance, the metric is called PGI (I for important masked first); in ascending order of importance, it is called PGU (U for unimportant masked first).
Fig.~\ref{fig:PG_scheme} shows a sketch of what to expect of these metrics. Note that the transition between the original prediction of the data sample and the prediction of the baseline can be less smooth. If the rankings are good, PGU estimates the fraction of irrelevant features, while PGI calculates the fraction of important features that are not necessary for the models.

\subsection{Assessing the genes selection methods}
\label{sec:comparison}
As the different selection methods generate distinct gene expression profiles, it is interesting to compare them. Rankings from classifier-based methods derive from learning, whose stability depends on the optimisation process. To account for variability, the ML models (LR, MLP, GNN) are trained 10 times with different initialisations, which generates 10 ranking replicates. In contrast, VAR, PCA, MI, EdgeR, and DESeq2 each generate a unique ranking per dataset.

\subsubsection{Comparing the top-ranked genes across methods}
For each dataset, heatmaps depicting the similarity of gene sets obtained by different methods are displayed (e.g., in Fig.~\ref{fig:heatmaps}). The lower / upper triangular part of the heatmap displays the percentages of common top 10 / 100 genes across different methods. When there are several ranking replicates, the percentages are averaged over all possible pairs. The diagonal of the heatmap shows the average percentage of common genes among the top 100 genes coming from ranking replicates of a method. For methods with a unique ranking, this percentage is 100\%, by design. 

\begin{table*}[t]
\centering
\begin{tabular}{|c|c|c|c|c|}
\hline
Dataset (source) & Task & \# classes & \# samples (min/max per class) & \# genes\\
\hline
PanCan (legacy TCGA) & Tumour types & 33 & 9853 (36/1095) & 15401 \\
BRCA (GDC TCGA) & Healthy vs Tumour & 2 & 1210 (113/1097) & 13946 \\
\small{BRCA-pam} (legacy TCGA) & PAM50 classes & 5 & 916 (67/421) & 13896 \\
ttg-breast (TCGA TARGET GTEx) & Healthy vs Tumour & 2 & 1384 (292/1092) & 14373\\
ttg-all (TCGA TARGET GTEx) & Healthy vs Tumour & 2 & 17600 (8130/9470) & 14368\\
\hline
\end{tabular}
\caption{Description of the various datasets and their associated classification tasks.}
\label{tab:data_description}
\end{table*}

\subsubsection{Classification performance investigation}
The analysis investigates the minimum number of genes needed for a model to achieve performance similar to using all genes. Two types of experiments are conducted. In \textit{experiment~1}, each model is trained using only the $n_H$ highest-ranked genes, with $n_H$ from 1 to 1000. In \textit{experiment~2}, each model is trained using the $n_L$ lowest-ranked genes, $n_L$ from 1 to 1000. For a model that produces several ranking replicates, the performance is averaged across all replicates.
These experiments aim to determine whether (1) the $n_H$ top-ranked genes contain enough information to discriminate the classes, and (2) whether the $n_L$ lowest-ranked genes are similarly informative.

\subsection{Biological function investigation}
After identifying the 100 most important genes for each method, an over-representation analysis (ORA)~\cite{khatri2012ten} is performed using gene sets from the Molecular Signatures Database Human collections~\cite{liberzon2011molecular}\footnote{\footnotesize{www.gsea-msigdb.org/gsea/msigdb/human/annotate.jsp}}. These gene sets, derived from the analysis of large repositories of gene expression data, encompass a variety of biological processes and pathological conditions. Specifically, the overlap of each gene list is calculated with the H hallmark gene sets~\cite{liberzon2015molecular}, the C2 curated gene sets (including canonical pathways), the C4 computational gene sets (curated from cancer expression data), the C5 ontology gene sets, the C6 oncogenic signature gene sets and the C7 immunologic signature gene sets. Results are filtered according to significance ($\text{false discovery rate q-value} \leq 0.05$). For each method, the top 10 over-represented gene sets are displayed. If any of these gene sets appears among the top 100 significant gene sets identified by another method, it is flagged. For ML-based methods, only one ranking is considered.

\subsection{Materials for the experiments}
\subsubsection{Datasets}

Bulk RNA-sequencing (RNA-seq) is a powerful technique for quantifying gene expression levels in tissue samples. The datasets analysed in this study originate from the TCGA Research Network,
%\footnote{www.cancer.gov/tcga}
the TARGET initiative
%\footnote{www.cancer.gov/ccg/research/genome-sequencing/target}
and the GTEx project~\cite{lonsdale2013genotype}. They are accessible through the Genomic Data Commons~\cite{grossman2016toward}\footnote{\footnotesize{portal.gdc.cancer.gov}} or the Xena browser~\cite{goldman2020visualizing}.
TCGA~\cite{tomczak2015review} encompasses gene expression data from various human tumours and surrounding normal tissues, available in legacy and Genomic Data Commons (GDC) versions. Notable distinctions between the two versions include the use of different human reference genomes (hg19 for legacy, hg38 for GDC)~\cite{gao2019before}. The Pan-Cancer Atlas, derived from legacy TCGA, explores 33 tumour types~\cite{hoadley2018cell}. The TCGA TARGET GTEx dataset (TTG) integrates gene expression data from adult cancer tissues (TCGA), pediatric cancer tissues (TARGET), and non-diseased tissues (GTEx), all processed using a unified pipeline~\cite{vivian2017toil}.

This article focuses on classical classification problems, such as PanCan~\cite{ramirez2020classification} or BRCA-pam~\cite{rhee2018hybrid, choi2023mobrca}. The datasets, detailed in Table~\ref{tab:data_description}, involve a limited number of samples for a large number of features (gene expression). 
Gene expression is measured by counting reads (fragment of RNA molecules) in a data sample.
Preprocessing steps include the removal of genes with NaN values, those exhibiting a maximal expression level of 0 across the dataset, and low-expressed genes with fewer than 5 counts in over 75\% of samples for each class. In the case of ttg-all, 27 samples are excluded due to null expression levels in more than 75\% of the genes. Normalisation involves scaling total counts in each sample to $10^6$, with subsequent $\text{log}_2$ transformation of the normalised counts. All datasets are available for download from our GitHub repository.

\subsubsection{Hyperparameter selection for ML methods}
The data is split randomly with 60\% of the samples used for training models and 40\% for testing. The hyperparameters of the models are selected with a grid search using a 4-split cross-validation on the training data. The obtained parameters are in Table~\ref{tab:data_hyperparameter}. For the graph, the correlation threshold is set such that the number of edges is $k \times G$, hence controlling its density. 

\begin{table}[t]
\centering
\begin{tabular}{|c|c|c|c|c|}
\hline
Dataset & LR+L1 & LR+L2 & MLP & GNN \\
\hline
PanCan & $95.0$ & $94.3$ & $94.3 \pm 0.3$ & $92.1 \pm 0.4$\\
BRCA & $99.7$ & $98.5$ & $99.5 \pm 0.4$ & $98.9 \pm 0.6$\\
BRCA-pam & $92.3$ & $90.7 \pm 0.2$ & $87.4 \pm 1.8$ & $87.1 \pm 1.4$\\
ttg-breast & $99.7$ & $99.2$ & $99.4 \pm 0.3$ & $99.1 \pm 0.1$\\
ttg-all & $99.5$ & $99.5$ & $99.6$ & $99.4 \pm 0.1$\\
\hline
\end{tabular}

\caption{Classification performance measured by balanced accuracy (\%). Standard deviations are computed from 10 replicates. They are not reported when below 0.05.}
\label{tab:classif_perf}
\end{table}

\subsubsection{Training process for ML methods}
Parameters of the models are learned through gradient descent, by minimising the cross-entropy loss function (plus regularisation). LR+L2 and LR+L1 are trained with the SAGA solver from~\cite{scikit-learn} with a maximum of 1000 iterations. MLP is trained for 25 epochs using PyTorch's SGD optimiser~\cite{paszke2019pytorch} with a momentum of 0.9 and a weight decay of 0.0001. The initial learning rate is 0.1, decreasing to 0.01 after 13 epochs and to 0.001 after 23 epochs. GNN is trained for 15 epochs using PyTorch's Adam optimiser~\cite{DBLP:journals/corr/KingmaB14} with a weight decay of 0.0001. The initial learning rate of 0.01 decreases to 0.001 after 8 epochs and to 0.0001 after 14 epochs. To address training variability, the learning process for each model is repeated ten times, using distinct initialisation seeds. This results in ten different rankings per model.

\subsubsection{Feature selection methods}
Gene rankings are generated using training samples only. All methods are implemented in Python except DESeq2 and EdgeR which are based on R packages. VAR, PCA and MI are implemented using scikit-learn~\cite{scikit-learn}. The code to include DESeq2 and EdgeR into a python script is inspired from~\cite{clarke2021appyters}. IG is implemented using the torch package called captum~\cite{kokhlikyan2020captum}. The IG scores are computed on the tumour class for BRCA, ttg-breast and ttg-all and on the tumour subtype classes for BRCA-pam. In these cases, the baseline $\vect{x}^\prime$ with respect to which the scores are computed is the average of the normal training samples. The scores are computed on all classes for PanCan with respect to the average of the training samples.

\section{Results}
\subsection{Gene expression data is informative on phenotypes}

\begin{table}[t]
\centering
\begin{tabular}{|c|c|c|c|c|}
\hline
Dataset & LR (non-zero for LR+L1) & MLP & GNN\\
\hline
PanCan & 508266 (13575) & 308773 & 969075\\
BRCA & 13947 (99) & 279001 & 26187\\
BRCA-pam & 69485 (390) & 278085 & 118484\\
ttg-breast & 14374 (206) & 287541 & 12501\\
ttg-all & 14369 (7376) & 576601 & 27456\\
\hline
\end{tabular}

\caption{Numbers of parameters learned by each model (averaged for selected non-zero parameters for LR+L1, and for GNN).}
\label{tab:classif_number_parameters}
\end{table}

Four ML models, LR+L1, LR+L2, MLP and GNN, are trained to classify gene expression data from tissues across several cancer types (PanCan), breast cancer and healthy surrounding tissues (BRCA), breast cancer and healthy tissues (ttg-breast), several cancer types and healthy tissues (ttg-all) and various subtypes of breast cancer (BRCA-pam) (Table~\ref{tab:data_description}). The classification performance is evaluated on data samples that have not been seen during training. Each model is trained 10 times with a different random initialisation. Table~\ref{tab:classif_perf} reports the average balanced accuracy scores, correcting for class imbalance (average of recall obtained on each class).  
Unbalanced accuracies are similar, see Table~\ref{tab:classif_perf2}. The numbers of parameters for the models are  in Table~\ref{tab:classif_number_parameters}.

Across all datasets and models, phenotypes are predicted with balanced accuracy consistently exceeding 95\% (except for BRCA-pam). High accuracies exceeding 99\% are even achieved for datasets classifying cancer tissues against normal tissues. For cancer types and subtypes, the best accuracies are respectively 95\% and 92\%. LR+L1 often outperforms other models (at the price of larger training time, see Table~\ref{tab:classif_time2}), with LR+L2 and MLP following closely. 

Good performance suggests also that the models have learned informative patterns linking gene expression data to phenotypes. As LR+L1 performed similarly to MLPs and GNNs, it suggests that precise classification can be achieved without necessarily learning interactions between genes; GNNs do not appear to benefit much from the graph of correlations.

\subsection{Top-ranked genes obtained by explainability and statistical gene selection methods differ significantly}
\begin{figure}[t]
\centering
\input{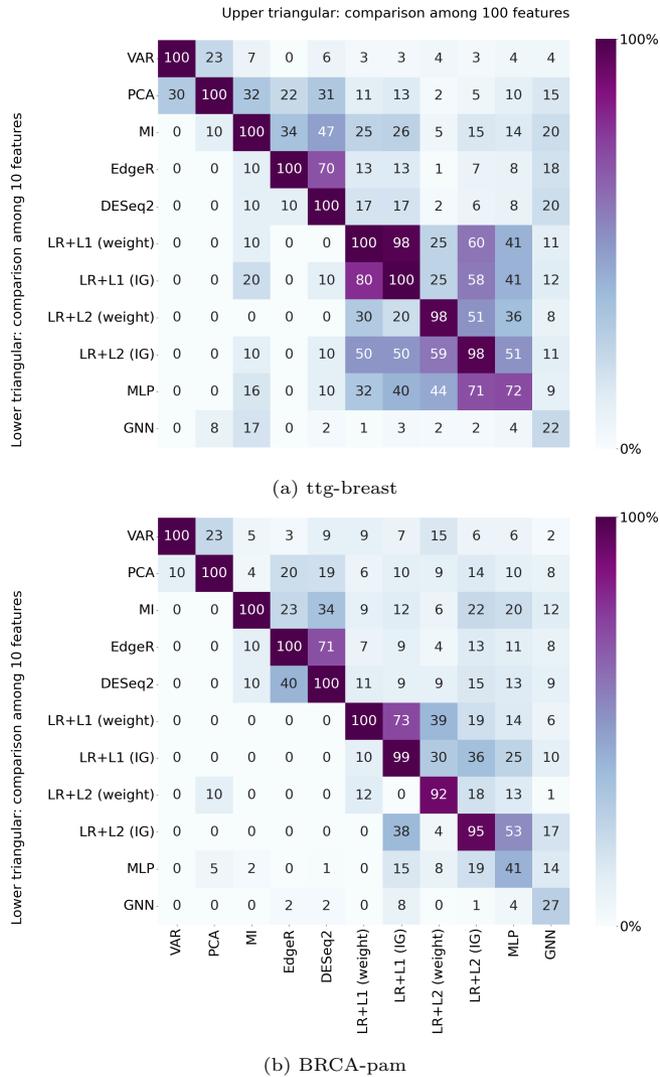}
\caption{Heatmaps showing the percentage of common genes among the top 10 (lower) and top 100 (upper + diagonal) genes selected by each method. More details are in Materials and methods.}
\label{fig:heatmaps}
\end{figure}

For each dataset, genes are ranked by the methods associated with the ML models (LR, MLP, GNN) or the statistical filtering methods (VAR, PCA, MI, EdgeR, DESeq2). Heatmaps are constructed to visualise the similarity of the top-ranked genes (Fig.~\ref{fig:heatmaps} for datasets ttg-breast and BRCA-pam and Fig.~\ref{fig:heatmaps2} for the other datasets).

For ttg-breast, LR exhibits more stable ranking replicates than MLP and GNN (diagonal of the heatmap). As expected, the more constant the classification performance (Table~\ref{tab:classif_perf}), the more similar the ranking between replicates. Across methods, the top-ranked genes may differ significantly. Classifier-based methods, except GNN, tend to select similar genes, possibly reflecting linear relationships with the class. LR with the embedded (weight) and post-hoc (IG) scores generate distinct rankings, especially for LR+L2. 
As IG is also applied to MLP and GNN, results across ML models are often more consistent when LR (IG) is considered instead of LR (weight). Thus, in the following, experiments are only conducted with IG. 
For the other datasets, qualitatively similar results are observed, with even greater differences between the selected gene sets obtained by ML models and statistical methods.

\subsection{Explaining the predictions with a small set of genes}
\begin{figure}[t]
\centering
\begin{tikzpicture}
\begin{groupplot}[group style={group size=2 by 2, horizontal sep=0.2cm, vertical sep=0.4cm}]

    \nextgroupplot[width=0.3\textwidth, 
        height=0.1\textwidth,
        hide axis,
	    ]

    \nextgroupplot[width=0.3\textwidth, 
        height=0.1\textwidth,
        xmin=0,
        xmax=100,
        ymin=0,
        ymax=100,
        legend columns=-1,
        legend style={at={(0.5,1)},anchor=east, fill opacity=0.5, text opacity=1},
        hide axis,
	    ]

     \addlegendimage{jublue1, thick, mark=o}
     \addlegendentry{PGU}
     \addlegendimage{orange, thick, mark=o}
     \addlegendentry{100 - PGI}

    \nextgroupplot[
        width=0.29\textwidth, 
        height=0.19\textwidth,
        ylabel style={align=center},
        xlabel style={align=center},
        xlabel={\small{Proportion of genes (\%)}\\ (a) ttg-breast},
        yticklabels={LR+L1, LR+L2, MLP, GNN},
        ytick={1,2,3,4},
        % xmajorgrids,
        ylabel near ticks,
	  xmax=55,
        xmin=-5,
	    ]

     \addplot[orange, thick, only marks, mark=o, error bars/.cd, x dir=both, x explicit] table [y=Index, x expr=100 -\thisrow{Value}, col sep=comma, x error=Std]{csv/evaluate/ttg-breast/PGI_global_train.csv};  % only marks

     \addplot[jublue1, thick, only marks, mark=o, error bars/.cd, x dir=both, x explicit] table [y=Index, x=Value, col sep=comma, x error=Std]{csv/evaluate/ttg-breast/PGU_global_train.csv};

    \nextgroupplot[
        width=0.29\textwidth, 
        height=0.19\textwidth,
        ylabel style={align=center},
        xlabel style={align=center},
        xlabel={\small{Proportion of genes (\%)}\\(b) BRCA-pam},
        yticklabels={, , , , }, %LR+L1, LR+L2, MLP, GNN},
        ytick={1,2,3,4},
        % xmajorgrids,
        ylabel near ticks,
        %ylabel=Values,
	  xmax=55,
        xmin=-5,
	    ]

     \addplot[orange, thick, only marks, mark=o, error bars/.cd, x dir=both, x explicit] table [y=Index, x expr=100 -\thisrow{Value}, col sep=comma, x error=Std]{csv/evaluate/BRCA-pam/PGI_global_train.csv};  % only marks 

     \addplot[jublue1, thick, only marks, mark=o, error bars/.cd, x dir=both, x explicit] table [y=Index, x=Value, col sep=comma, x error=Std]{csv/evaluate/BRCA-pam/PGU_global_train.csv};

\end{groupplot}
\end{tikzpicture}
\vspace{-0.5cm}
\caption{Impact of progressive gene masking on the predictions of ML models (\textit{experiment~0}). Genes are masked by increasing (PGU) or decreasing order of importance (PGI) based on the rankings $\boldsymbol{\phi}^\text{IG}$. 
For each data sample, PGU calculates the percentage of well-ranked genes that should remain unmasked to avoid disturbing a trained model. $100 -$PGI estimates the percentage of well-ranked genes that can be masked before disturbing the model. PGs are averaged over all training samples correctly classified. Error bars are standard deviations across replicates.
}
\label{fig:PGs}
\end{figure}
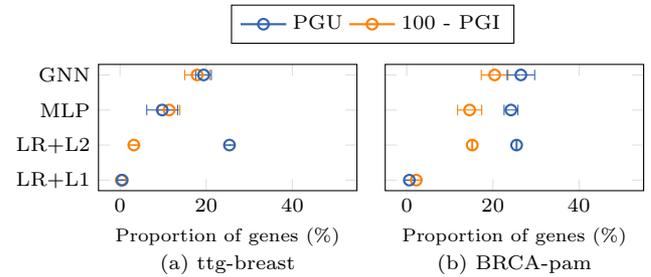

We investigate the minimum number of genes needed for a model to attain performance comparable to using all genes. 
With LR+L1, the optimisation process promotes sparsity, leading to many null parameters and the selection of a small number of crucial genes, as shown in Table~\ref{tab:classif_number_parameters}. The precise number depends heavily on the choice of the hyperparameter $\lambda$.
Still, the overall strong performance of LR+L1 indicates the ability to classify phenotypes effectively in these datasets with a limited number of genes.

Experiments are conducted to explore ML models.
\textit{Experiment~0} is focused on the ML-based rankings derived from IG. The number of genes used by a trained classifier to make a decision is estimated by progressively masking the lowest-ranked genes without re-training. 
\textit{Experiment~1} measures the performance of models when they are trained from scratch using only the top-ranked genes identified by IG and statistical methods. The goal is to assess whether the top-ranked genes is sufficient for classification.

For \textit{experiment~0}, in Fig.~\ref{fig:PGs}, the Prediction Gap on Unimportant features (PGU) measures the minimum percentage of top-ranked genes that must remain unmasked to avoid disturbing the model's decision. For ttg-breast, PGUs for LR+L1 is about 0.5\% of genes (around 70 genes), which is less than the number of non-zero parameters in this model (Table~\ref{tab:classif_number_parameters}). PGUs for MLP, GNN and LR+L2 are 10\%, 19\% and 25\% of genes respectively (around 1400, 2700 and 3600 genes). For BRCA-pam, LR+L1 needs 0.6\% of genes (around 80 genes), while other models have to keep a larger number of genes, between 24\% and 27\% (3300 to 3800 genes). Results on the other datasets are shown in Fig.~\ref{fig:PGs2}.

\begin{figure*}[h]
\centering
\input{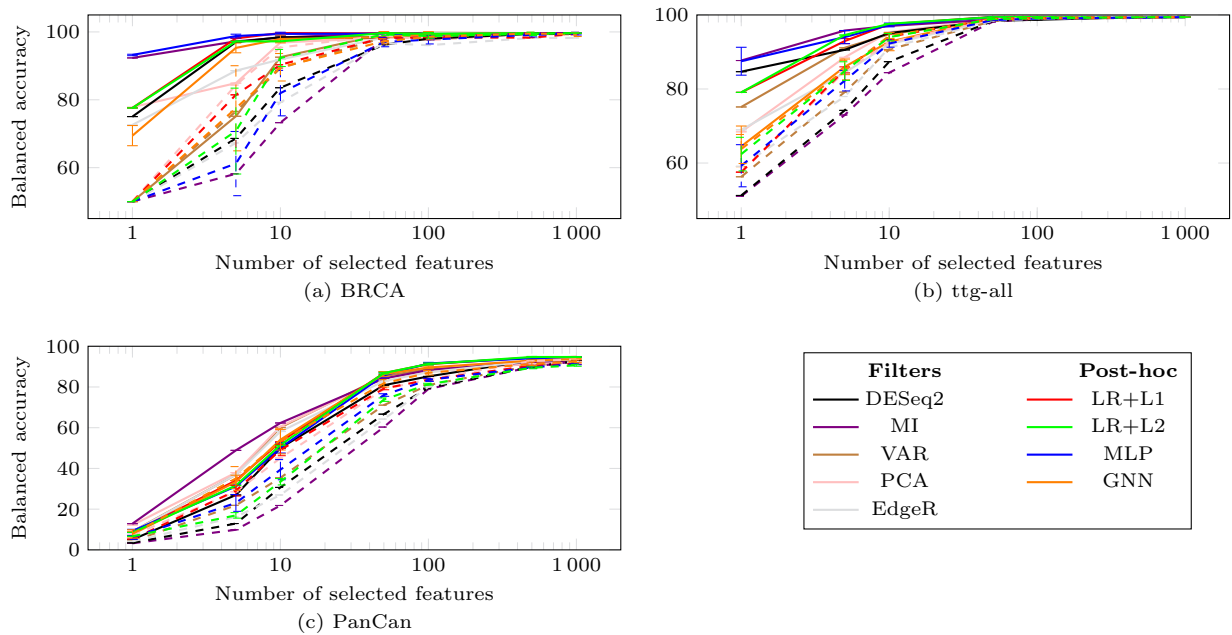}
\caption{Classification performance of a MLP trained on sets of features identified as important by various methods.}
\label{fig:classif_perf_MLP2}
\end{figure*}

For \textit{experiment~1}, accuracies computed after re-training the models are in Fig.~\ref{fig:classif_perf}. For ttg-breast, the best 10 (resp.~100) genes are sufficient for MLP, LR+L1 and LR+L2 (resp.~GNN) to saturate the classification performance. For BRCA-pam, 500 genes are sufficient for all models. On both datasets, LRs and MLPs obtain better performance than GNNs, which are disadvantaged because the graph structure is highly perturbed. Note that better classification performance can be obtained for MLP when trained on gene sets selected by other methods (Fig.~\ref{fig:classif_perf_MLP}). Results on the other datasets are in Figs.~\ref{fig:classif_perf2} and~\ref{fig:classif_perf_MLP2}.

Thus, for a model trained on all genes, most genes but a very small sets of the \emph{best} genes can be masked without significantly perturbing predictions. Even smaller sets of the \emph{best} genes are sufficient to train models with high performance.

\subsection{{\it Unimportant} genes matter}

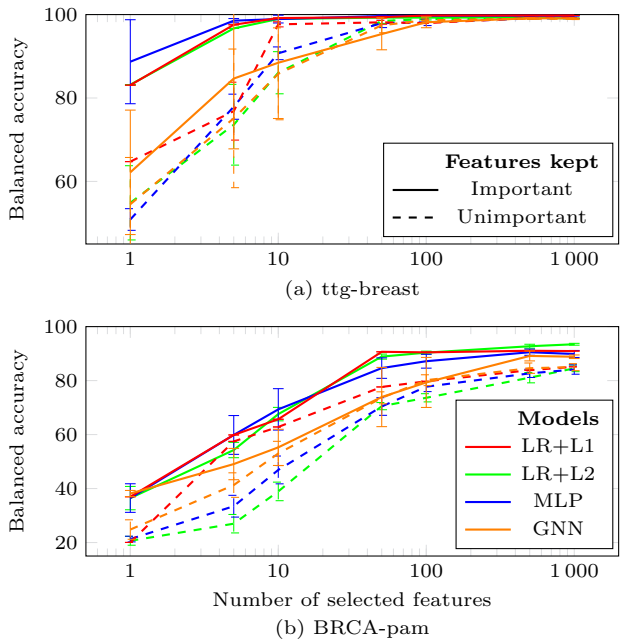
\begin{figure}[t]
\centering
\begin{tikzpicture}
\begin{groupplot}[group style={group size=1 by 3, horizontal sep=1cm, vertical sep=1.1cm}]
    
    \nextgroupplot[
        width=0.5\textwidth, 
        height=0.27\textwidth,
        ylabel style={align=center},
        xlabel style={align=center},
        xlabel={(a) ttg-breast},
        ylabel=Balanced accuracy,
        ymax=100,
        ymin=45,
        xmode=log,
        log ticks with fixed point,
        legend style={at={(0.99, 0.03)}, anchor=south east,}
	    ]

    % Legend
    \addlegendimage{empty legend}
    \addlegendentry{\textbf{Features kept}}

    \addlegendimage{black, thick}
    \addlegendentry{Important}

    \addlegendimage{black, dashed, thick}
    \addlegendentry{Unimportant}

    % LR L2
     \addplot[green, thick, mark=none, error bars/.cd, y dir=both, y explicit] table [x=Features, y=Important, col sep=comma, y error=Std_important]{csv/compare/self/ttg_breast_LR_L2.csv};
     \addplot[green, dashed, thick, mark=none, error bars/.cd, y dir=both, y explicit] table [x=Features, y=Unimportant, col sep=comma, y error=Std_unimportant]{csv/compare/self/ttg_breast_LR_L2.csv};

     % MLP
     \addplot[blue, thick, mark=none, error bars/.cd, y dir=both, y explicit] table [x=Features, y=Important, col sep=comma, y error=Std_important]{csv/compare/self/ttg_breast_MLP.csv};
     \addplot[blue, dashed, thick, mark=none, error bars/.cd, y dir=both, y explicit] table [x=Features, y=Unimportant, col sep=comma, y error=Std_unimportant]{csv/compare/self/ttg_breast_MLP.csv};

    % LR L1
     \addplot[red, thick, mark=none, error bars/.cd, y dir=both, y explicit] table [x=Features, y=Important, col sep=comma, y error=Std_important]{csv/compare/self/ttg_breast_LR_L1.csv};
     \addplot[red, dashed, thick, mark=none, error bars/.cd, y dir=both, y explicit] table [x=Features, y=Unimportant, col sep=comma, y error=Std_unimportant]{csv/compare/self/ttg_breast_LR_L1.csv};

     % GCN
     \addplot[orange, thick, mark=none, error bars/.cd, y dir=both, y explicit] table [x=Features, y=Important, col sep=comma, y error=Std_important]{csv/compare/self/ttg_breast_GCN.csv};
     \addplot[orange, dashed, thick, mark=none, error bars/.cd, y dir=both, y explicit] table [x=Features, y=Unimportant, col sep=comma, y error=Std_unimportant]{csv/compare/self/ttg_breast_GCN.csv};

    \nextgroupplot[
        width=0.5\textwidth, 
        height=0.27\textwidth,
        ylabel style={align=center},
        xlabel style={align=center},
        xlabel={Number of selected features\\(b) BRCA-pam},
        ylabel=Balanced accuracy,
        ymax=100,
        ymin=15,
        xmode=log,
        log ticks with fixed point,
        legend style={at={(0.99, 0.03)}, anchor=south east,}
	    ]

    % Legend
    \addlegendimage{empty legend}
    \addlegendentry{\textbf{Models}}
    
    \addlegendimage{red, thick}
    \addlegendentry{LR+L1}
    
    \addlegendimage{green, thick}
    \addlegendentry{LR+L2} 
    
    \addlegendimage{blue, thick}
    \addlegendentry{MLP}
    
    \addlegendimage{orange, thick}
    \addlegendentry{GNN} 
    
    % LR L2
     \addplot[green, thick, mark=none, error bars/.cd, y dir=both, y explicit] table [x=Features, y=Important, col sep=comma, y error=Std_important]{csv/compare/self/BRCA_pam_LR_L2.csv};
     \addplot[green, dashed, thick, mark=none, error bars/.cd, y dir=both, y explicit] table [x=Features, y=Unimportant, col sep=comma, y error=Std_unimportant]{csv/compare/self/BRCA_pam_LR_L2.csv};

     % MLP
     \addplot[blue, thick, mark=none, error bars/.cd, y dir=both, y explicit] table [x=Features, y=Important, col sep=comma, y error=Std_important]{csv/compare/self/BRCA_pam_MLP.csv};
     \addplot[blue, dashed, thick, mark=none, error bars/.cd, y dir=both, y explicit] table [x=Features, y=Unimportant, col sep=comma, y error=Std_unimportant]{csv/compare/self/BRCA_pam_MLP.csv};

    % LR L1
     \addplot[red, thick, mark=none, error bars/.cd, y dir=both, y explicit] table [x=Features, y=Important, col sep=comma, y error=Std_important]{csv/compare/self/BRCA_pam_LR_L1.csv};
     \addplot[red, dashed, thick, mark=none, error bars/.cd, y dir=both, y explicit] table [x=Features, y=Unimportant, col sep=comma, y error=Std_unimportant]{csv/compare/self/BRCA_pam_LR_L1.csv};

     % GCN
     \addplot[orange, thick, mark=none, error bars/.cd, y dir=both, y explicit] table [x=Features, y=Important, col sep=comma, y error=Std_important]{csv/compare/self/BRCA_pam_GCN.csv};
     \addplot[orange, dashed, thick, mark=none, error bars/.cd, y dir=both, y explicit] table [x=Features, y=Unimportant, col sep=comma, y error=Std_unimportant]{csv/compare/self/BRCA_pam_GCN.csv};

\end{groupplot}
\end{tikzpicture}
\caption{Classification performance shown for models trained on features identified as important (full lines, \textit{experiment~1}) or unimportant (dashed lines, \textit{experiment~2}). Balanced accuracies are reported as a function of the number of kept features for ttg-breast (a) and BRCA-pam (b) datasets using the specified models. Error bars are std from 10 replicates.}
\label{fig:classif_perf}
\end{figure}

This analysis investigates whether the genes selected in the previous section exclusively contain relevant information. For that, we conduct \textit{experiment~0} by masking the highest-ranked genes without re-training the models and \textit{experiment~2} for which models are trained using the $n_L$ lowest-ranked genes.

For \textit{experiment~0}, in Fig.~\ref{fig:PGs}, the Prediction Gap on Important features (PGI) measures the minimum percentage of worst-ranked genes that must remain unmasked to avoid disturbing the model's decision. Here, the orange dots represent $100 -$PGI values, indicating the highest percentage of well-ranked genes that can be masked before perturbing the model. When the orange dots surpass the blue dots (PGU), it suggests that all the genes selected as important in the previous subsection can be masked without disturbing the model. Thus, the remaining information is sufficiently redundant. For ttg-breast, this occurs for MLP only. For LR+L2, masking a small proportion of the identified important genes is enough to disrupt the model. For BRCA-pam, LR+L1 can mask all the identified important elements. In contrast, for MLP, LR+L2, and GNN, masking a small proportion of the important genes is sufficient. These results show that when a large number of genes is perturbed, the model is disrupted. However, when the model is based on a small number of genes, these genes are not necessarily the only ones containing relevant information. Results on the other datasets are in Fig.~\ref{fig:PGs2}.

\begin{figure*}[h]
\centering
\begin{tikzpicture}
\begin{groupplot}[group style={group size=2 by 3, horizontal sep=1cm, vertical sep=1.7cm}]

    \nextgroupplot[
        width=0.5\textwidth, 
        height=0.25\textwidth,
        ylabel style={align=center},
        xlabel style={align=center},
        xlabel={Number of selected features\\(a) BRCA},
        ylabel=Balanced accuracy,
        ymax=105,
        ymin=45,
        xmode=log,
        log ticks with fixed point,
	    ]

    % LR L2
     \addplot[green, thick, mark=none, error bars/.cd, y dir=both, y explicit] table [x=Features, y=Important, col sep=comma, y error=Std_important]{csv/compare/self/BRCA_LR_L2.csv};
     \addplot[green, dashed, thick, mark=none, error bars/.cd, y dir=both, y explicit] table [x=Features, y=Unimportant, col sep=comma, y error=Std_unimportant]{csv/compare/self/BRCA_LR_L2.csv};

     % MLP
     \addplot[blue, thick, mark=none, error bars/.cd, y dir=both, y explicit] table [x=Features, y=Important, col sep=comma, y error=Std_important]{csv/compare/self/BRCA_MLP.csv};
     \addplot[blue, dashed, thick, mark=none, error bars/.cd, y dir=both, y explicit] table [x=Features, y=Unimportant, col sep=comma, y error=Std_unimportant]{csv/compare/self/BRCA_MLP.csv};

    % LR L1
     \addplot[red, thick, mark=none, error bars/.cd, y dir=both, y explicit] table [x=Features, y=Important, col sep=comma, y error=Std_important]{csv/compare/self/BRCA_LR_L1.csv};  
     \addplot[red, dashed, thick, mark=none, error bars/.cd, y dir=both, y explicit] table [x=Features, y=Unimportant, col sep=comma, y error=Std_unimportant]{csv/compare/self/BRCA_LR_L1.csv};

     % GCN
     \addplot[orange, thick, mark=none, error bars/.cd, y dir=both, y explicit] table [x=Features, y=Important, col sep=comma, y error=Std_important]{csv/compare/self/BRCA_GCN.csv};
     \addplot[orange, dashed, thick, mark=none, error bars/.cd, y dir=both, y explicit] table [x=Features, y=Unimportant, col sep=comma, y error=Std_unimportant]{csv/compare/self/BRCA_GCN.csv};

    \nextgroupplot[
        width=0.5\textwidth, 
        height=0.25\textwidth,
        ylabel style={align=center},
        xlabel style={align=center},
        xlabel={Number of selected features\\(b) ttg-all},
        ymax=100,
        ymin=45,
        xmode=log,
        log ticks with fixed point,
	    ]

    % LR L2
     \addplot[green, thick, mark=none, error bars/.cd, y dir=both, y explicit] table [x=Features, y=Important, col sep=comma, y error=Std_important]{csv/compare/self/ttg_all_LR_L2.csv};
     \addplot[green, dashed, thick, mark=none, error bars/.cd, y dir=both, y explicit] table [x=Features, y=Unimportant, col sep=comma, y error=Std_unimportant]{csv/compare/self/ttg_all_LR_L2.csv};

     % MLP
     \addplot[blue, thick, mark=none, error bars/.cd, y dir=both, y explicit] table [x=Features, y=Important, col sep=comma, y error=Std_important]{csv/compare/self/ttg_all_MLP.csv};
     \addplot[blue, dashed, thick, mark=none, error bars/.cd, y dir=both, y explicit] table [x=Features, y=Unimportant, col sep=comma, y error=Std_unimportant]{csv/compare/self/ttg_all_MLP.csv};

    % LR L1
     \addplot[red, thick, mark=none, error bars/.cd, y dir=both, y explicit] table [x=Features, y=Important, col sep=comma, y error=Std_important]{csv/compare/self/ttg_all_LR_L1.csv};
     \addplot[red, dashed, thick, mark=none, error bars/.cd, y dir=both, y explicit] table [x=Features, y=Unimportant, col sep=comma, y error=Std_unimportant]{csv/compare/self/ttg_all_LR_L1.csv};

     % GCN
     \addplot[orange, thick, mark=none, error bars/.cd, y dir=both, y explicit] table [x=Features, y=Important, col sep=comma, y error=Std_important]{csv/compare/self/ttg_all_GCN.csv};
     \addplot[orange, dashed, thick, mark=none, error bars/.cd, y dir=both, y explicit] table [x=Features, y=Unimportant, col sep=comma, y error=Std_unimportant]{csv/compare/self/ttg_all_GCN.csv};

    \nextgroupplot[
        width=0.5\textwidth, 
        height=0.25\textwidth,
        ylabel style={align=center},
        xlabel style={align=center},
        xlabel={Number of selected features\\(c) PanCan},
        ylabel=Balanced accuracy,
        ymax=100,
        ymin=0,
        xmode=log,
        log ticks with fixed point,
	    ]

    % LR L2
     \addplot[green, thick, mark=none, error bars/.cd, y dir=both, y explicit] table [x=Features, y=Important, col sep=comma, y error=Std_important]{csv/compare/self/pancan_LR_L2.csv};
     \addplot[green, dashed, thick, mark=none, error bars/.cd, y dir=both, y explicit] table [x=Features, y=Unimportant, col sep=comma, y error=Std_unimportant]{csv/compare/self/pancan_LR_L2.csv};

     % MLP
     \addplot[blue, thick, mark=none, error bars/.cd, y dir=both, y explicit] table [x=Features, y=Important, col sep=comma, y error=Std_important]{csv/compare/self/pancan_MLP.csv}; 
     \addplot[blue, dashed, thick, mark=none, error bars/.cd, y dir=both, y explicit] table [x=Features, y=Unimportant, col sep=comma, y error=Std_unimportant]{csv/compare/self/pancan_MLP.csv};

    % LR L1
     \addplot[red, thick, mark=none, error bars/.cd, y dir=both, y explicit] table [x=Features, y=Important, col sep=comma, y error=Std_important]{csv/compare/self/pancan_LR_L1.csv};
     \addplot[red, dashed, thick, mark=none, error bars/.cd, y dir=both, y explicit] table [x=Features, y=Unimportant, col sep=comma, y error=Std_unimportant]{csv/compare/self/pancan_LR_L1.csv};

     % GCN
     \addplot[orange, thick, mark=none, error bars/.cd, y dir=both, y explicit] table [x=Features, y=Important, col sep=comma, y error=Std_important]{csv/compare/self/pancan_GCN.csv};
     \addplot[orange, dashed, thick, mark=none, error bars/.cd, y dir=both, y explicit] table [x=Features, y=Unimportant, col sep=comma, y error=Std_unimportant]{csv/compare/self/pancan_GCN.csv};

    \nextgroupplot[
        width=0.5\textwidth, 
        height=0.25\textwidth,
        hide axis,
        xmin=0,
        xmax=100,
        ymin=0,
        ymax=100,
        legend columns=2,
        legend style={at={(0.1, 0.15)}, anchor=south west, /tikz/every even column/.append style={column sep=1cm}}
	    ]
    
        \addlegendimage{empty legend}
        \addlegendentry{\textbf{Features kept}}
        \addlegendimage{empty legend}
        \addlegendentry{\textbf{Models}}
        
        \addlegendimage{black, thick}
        \addlegendentry{Important}
        \addlegendimage{red, thick}
        \addlegendentry{LR+L1}
        
        \addlegendimage{black, dashed, thick}
        \addlegendentry{Unimportant}
        \addlegendimage{green, thick}
        \addlegendentry{LR+L2} 
        
        \addlegendimage{white}
        \addlegendentry{}
        \addlegendimage{blue, thick}
        \addlegendentry{MLP}
        
        \addlegendimage{white} 
        \addlegendentry{}
        \addlegendimage{orange, thick}
        \addlegendentry{GNN}

\end{groupplot}
\end{tikzpicture}
\caption{Classification performance shown for models trained on features identified as important or unimportant for each model. Balanced accuracies are reported as a function of the number of features kept using the specified models. Error bars are std from 10 replicates. 
\vspace*{-1cm}}
\label{fig:classif_perf2}
\end{figure*}
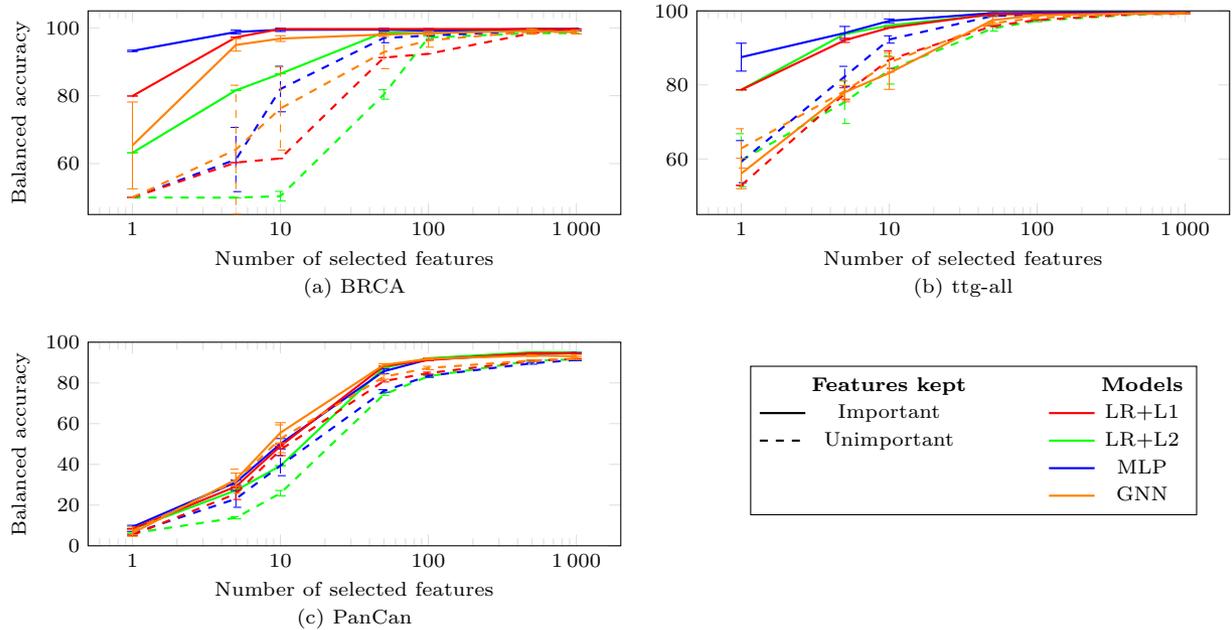

For \textit{experiment~2}, with re-training, results are in Fig.~\ref{fig:classif_perf} (dashed lines). Using the 100 lowest-ranked genes is sufficient to achieve optimal balanced accuracy on ttg-breast. For BRCA-pam, the gaps between the full lines (best-ranked genes) and the dashed lines (worst-ranked genes) are small. Results on the other datasets are in Fig.~\ref{fig:classif_perf2}.

Similar classification performance can be achieved by keeping a relatively small set of lower-ranked genes, showing there is no unique set of informative genes.

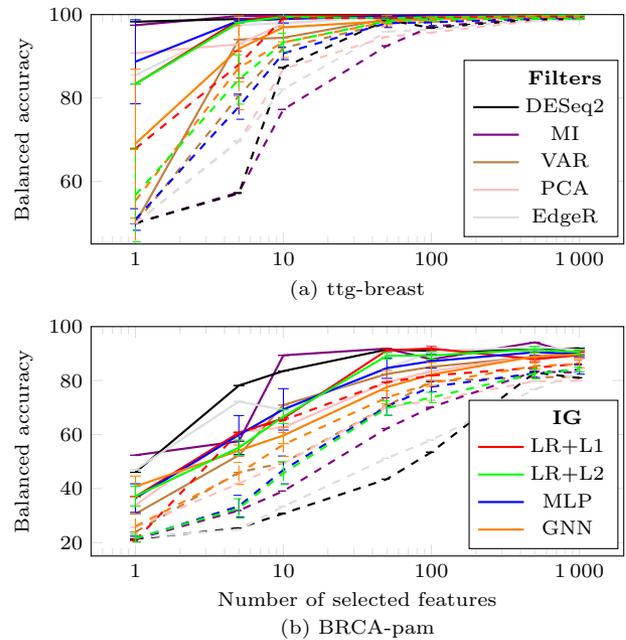
\begin{figure}
\centering
\begin{tikzpicture}
\begin{groupplot}[group style={group size=1 by 2, horizontal sep=1cm, vertical sep=1.1cm}]

    \nextgroupplot[
        width=0.5\textwidth, 
        height=0.27\textwidth,
        ylabel style={align=center},
        xlabel style={align=center},
        xlabel={(a) ttg-breast},
        ylabel=Balanced accuracy,
        ymax=100,
        ymin=45,
        xmode=log,
        log ticks with fixed point,
        legend style={at={(0.99, 0.03)}, anchor=south east,}
	    ]

    % Legend
    \addlegendimage{empty legend}
    \addlegendentry{\textbf{Filters}}

    \addlegendimage{black, thick}
    \addlegendentry{DESeq2}

    \addlegendimage{violet, thick}
    \addlegendentry{MI} 

    \addlegendimage{brown, thick}
    \addlegendentry{VAR}

    \addlegendimage{pink, thick}
    \addlegendentry{PCA}

    \addlegendimage{gray, thick}
    \addlegendentry{EdgeR}
    
    % Best
    % Filter
     \addplot[brown, thick, mark=none, error bars/.cd, y dir=both, y explicit] table [x=Features, y=VAR, col sep=comma, y error=STD_VAR]{csv/compare/other/best/ttg_breast_MLP.csv};
     \addplot[pink, thick, mark=none, error bars/.cd, y dir=both, y explicit] table [x=Features, y=PCA, col sep=comma, y error=STD_PCA]{csv/compare/other/best/ttg_breast_MLP.csv};
     \addplot[violet, thick, mark=none, error bars/.cd, y dir=both, y explicit] table [x=Features, y=MI, col sep=comma, y error=STD_MI]{csv/compare/other/best/ttg_breast_MLP.csv};
     \addplot[black, thick, mark=none, error bars/.cd, y dir=both, y explicit] table [x=Features, y=DESeq2, col sep=comma, y error=STD_DESeq2]{csv/compare/other/best/ttg_breast_MLP.csv};
     \addplot[gray, thick, mark=none, error bars/.cd, y dir=both, y explicit] table [x=Features, y=EdgeR, col sep=comma, y error=STD_EdgeR]{csv/compare/other/best/ttg_breast_MLP.csv};   

    % Post-hoc
    \addplot[red, thick, mark=none, error bars/.cd, y dir=both, y explicit] table [x=Features, y=LR_L1, col sep=comma, y error=STD_LR_L1]{csv/compare/other/best/ttg_breast_MLP.csv};
     \addplot[blue, thick, mark=none, error bars/.cd, y dir=both, y explicit] table [x=Features, y=MLP, col sep=comma, y error=STD_MLP]{csv/compare/other/best/ttg_breast_MLP.csv};
     \addplot[orange, thick, mark=none, error bars/.cd, y dir=both, y explicit] table [x=Features, y=GCN, col sep=comma, y error=STD_GCN]{csv/compare/other/best/ttg_breast_MLP.csv};

     % Self
     \addplot[green, thick, mark=none, error bars/.cd, y dir=both, y explicit] table [x=Features, y=LR_L2, col sep=comma, y error=STD_LR_L2]{csv/compare/other/best/ttg_breast_MLP.csv};

     % Worst
     % Filter
     \addplot[brown, thick, dashed, mark=none, error bars/.cd, y dir=both, y explicit] table [x=Features, y=VAR, col sep=comma, y error=STD_VAR]{csv/compare/other/worst/ttg_breast_MLP.csv};
     \addplot[pink, thick, dashed, mark=none, error bars/.cd, y dir=both, y explicit] table [x=Features, y=PCA, col sep=comma, y error=STD_PCA]{csv/compare/other/worst/ttg_breast_MLP.csv};
     \addplot[violet, thick, dashed, mark=none, error bars/.cd, y dir=both, y explicit] table [x=Features, y=MI, col sep=comma, y error=STD_MI]{csv/compare/other/worst/ttg_breast_MLP.csv};
     \addplot[black, thick, dashed, mark=none, error bars/.cd, y dir=both, y explicit] table [x=Features, y=DESeq2, col sep=comma, y error=STD_DESeq2]{csv/compare/other/worst/ttg_breast_MLP.csv};
     \addplot[gray, thick, dashed, mark=none, error bars/.cd, y dir=both, y explicit] table [x=Features, y=EdgeR, col sep=comma, y error=STD_EdgeR]{csv/compare/other/worst/ttg_breast_MLP.csv};

    % Post-hoc
    \addplot[red, thick, dashed, mark=none, error bars/.cd, y dir=both, y explicit] table [x=Features, y=LR_L1, col sep=comma, y error=STD_LR_L1]{csv/compare/other/worst/ttg_breast_MLP.csv};
     \addplot[blue, thick, dashed, mark=none, error bars/.cd, y dir=both, y explicit] table [x=Features, y=MLP, col sep=comma, y error=STD_MLP]{csv/compare/other/worst/ttg_breast_MLP.csv};
     \addplot[orange, thick, dashed, mark=none, error bars/.cd, y dir=both, y explicit] table [x=Features, y=GCN, col sep=comma, y error=STD_GCN]{csv/compare/other/worst/ttg_breast_MLP.csv};

     % Self
     \addplot[green, thick, dashed, mark=none, error bars/.cd, y dir=both, y explicit] table [x=Features, y=LR_L2, col sep=comma, y error=STD_LR_L2]{csv/compare/other/worst/ttg_breast_MLP.csv};

    \nextgroupplot[
        width=0.5\textwidth, 
        height=0.27\textwidth,
        ylabel style={align=center},
        xlabel style={align=center},
        xlabel={Number of selected features\\(b) BRCA-pam},
        % xmajorgrids,
        % xlabel near ticks,
        ylabel=Balanced accuracy,
        ymax=100,
        ymin=15,
        xmode=log,
        log ticks with fixed point,
        legend columns=1,
        legend style={at={(0.99, 0.03)}, anchor=south east}
	    ]

    % Legend
    \addlegendimage{empty legend}
    \addlegendentry{\textbf{IG}}

    \addlegendimage{red, thick}
    \addlegendentry{LR+L1}

    \addlegendimage{green, thick}
    \addlegendentry{LR+L2}

    \addlegendimage{blue, thick}
    \addlegendentry{MLP}
    
    \addlegendimage{orange, thick}
    \addlegendentry{GNN}

    % Best
    % Filter
     \addplot[brown, thick, mark=none, error bars/.cd, y dir=both, y explicit] table [x=Features, y=VAR, col sep=comma, y error=STD_VAR]{csv/compare/other/best/BRCA_pam_MLP.csv};
     \addplot[pink, thick, mark=none, error bars/.cd, y dir=both, y explicit] table [x=Features, y=PCA, col sep=comma, y error=STD_PCA]{csv/compare/other/best/BRCA_pam_MLP.csv};
     \addplot[violet, thick, mark=none, error bars/.cd, y dir=both, y explicit] table [x=Features, y=MI, col sep=comma, y error=STD_MI]{csv/compare/other/best/BRCA_pam_MLP.csv};
     \addplot[black, thick, mark=none, error bars/.cd, y dir=both, y explicit] table [x=Features, y=DESeq2, col sep=comma, y error=STD_DESeq2]{csv/compare/other/best/BRCA_pam_MLP.csv};
     \addplot[gray, thick, mark=none, error bars/.cd, y dir=both, y explicit] table [x=Features, y=EdgeR, col sep=comma, y error=STD_EdgeR]{csv/compare/other/best/BRCA_pam_MLP.csv};
     
    % Post-hoc
    \addplot[red, thick, mark=none, error bars/.cd, y dir=both, y explicit] table [x=Features, y=LR_L1, col sep=comma, y error=STD_LR_L1]{csv/compare/other/best/BRCA_pam_MLP.csv};
     \addplot[blue, thick, mark=none, error bars/.cd, y dir=both, y explicit] table [x=Features, y=MLP, col sep=comma, y error=STD_MLP]{csv/compare/other/best/BRCA_pam_MLP.csv};
     \addplot[orange, thick, mark=none, error bars/.cd, y dir=both, y explicit] table [x=Features, y=GCN, col sep=comma, y error=STD_GCN]{csv/compare/other/best/BRCA_pam_MLP.csv};

     % Self
     \addplot[green, thick, mark=none, error bars/.cd, y dir=both, y explicit] table [x=Features, y=LR_L2, col sep=comma, y error=STD_LR_L2]{csv/compare/other/best/BRCA_pam_MLP.csv};

     % Worst
     % Filter
     \addplot[brown, thick, dashed, mark=none, error bars/.cd, y dir=both, y explicit] table [x=Features, y=VAR, col sep=comma, y error=STD_VAR]{csv/compare/other/worst/BRCA_pam_MLP.csv};
     \addplot[pink, thick, dashed, mark=none, error bars/.cd, y dir=both, y explicit] table [x=Features, y=PCA, col sep=comma, y error=STD_PCA]{csv/compare/other/worst/BRCA_pam_MLP.csv};
     \addplot[violet, thick, dashed, mark=none, error bars/.cd, y dir=both, y explicit] table [x=Features, y=MI, col sep=comma, y error=STD_MI]{csv/compare/other/worst/BRCA_pam_MLP.csv};
     \addplot[black, thick, dashed, mark=none, error bars/.cd, y dir=both, y explicit] table [x=Features, y=DESeq2, col sep=comma, y error=STD_DESeq2]{csv/compare/other/worst/BRCA_pam_MLP.csv};
     \addplot[gray, thick, dashed, mark=none, error bars/.cd, y dir=both, y explicit] table [x=Features, y=EdgeR, col sep=comma, y error=STD_EdgeR]{csv/compare/other/worst/BRCA_pam_MLP.csv};

    % Post-hoc
    \addplot[red, thick, dashed, mark=none, error bars/.cd, y dir=both, y explicit] table [x=Features, y=LR_L1, col sep=comma, y error=STD_LR_L1]{csv/compare/other/worst/BRCA_pam_MLP.csv};
     \addplot[blue, thick, dashed, mark=none, error bars/.cd, y dir=both, y explicit] table [x=Features, y=MLP, col sep=comma, y error=STD_MLP]{csv/compare/other/worst/BRCA_pam_MLP.csv};
     \addplot[orange, thick, dashed, mark=none, error bars/.cd, y dir=both, y explicit] table [x=Features, y=GCN, col sep=comma, y error=STD_GCN]{csv/compare/other/worst/BRCA_pam_MLP.csv};

     % Self
     \addplot[green, thick, dashed, mark=none, error bars/.cd, y dir=both, y explicit] table [x=Features, y=LR_L2, col sep=comma, y error=STD_LR_L2]{csv/compare/other/worst/BRCA_pam_MLP.csv};

\end{groupplot}
\end{tikzpicture}
\caption{Classification performance of a MLP trained on sets of features identified as important by various methods as indicated. The representation is coded as in Fig.~\ref{fig:classif_perf}.}
\label{fig:classif_perf_MLP}
\end{figure}

\begin{figure*}[t]
\centering
\input{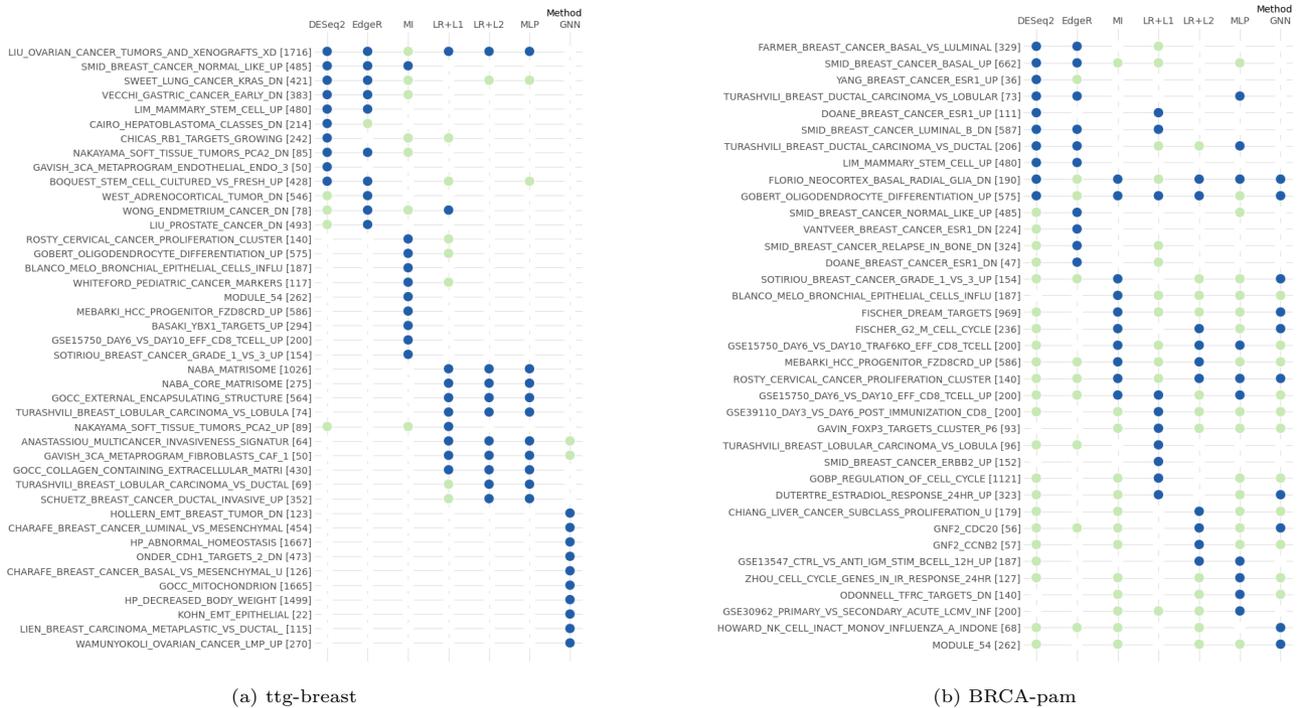}
\caption{Top 10 over-represented gene sets identified by the different methods (blue dots). When one of these gene sets appear among the top 100 over-represented sets of another method, it is flagged with a light green dot.}
\label{fig:GSEA}
\end{figure*}

\subsection{Stability across gene sets}
The top-ranked genes identified by the different methods are compared with the content of gene sets known for their biological properties using an over-representation analysis. 
For each method, the top 10 over-represented gene sets are displayed in Fig.~\ref{fig:GSEA} (blue dots). When these gene sets appear among the top 100 significant gene sets identified by another method, they are flagged with a light green dot. Interestingly, different sets associated with (breast) cancer emerge across different methods. However, the overlap between the methods can be small. See Fig.~\ref{fig:GSEA2} for the other datasets.

\section{Discussion}
% Goal
Many ML models have emerged to classify phenotypes from gene expression data, with the goal of profiling genes discriminative for the phenotypes. Model explanations are often presented as a list of genes ranked by importance. Our objective is to evaluate the relevance of explanations generated by ML models, examining whether they contribute novel perspectives to elucidate complex biological pathways.

% Methodology
To address this question, experiments are carried out on human tissue samples coming from TCGA, TARGET and GTEx databases. Three ML models, LRs, MLPs and GNNs, are trained to discern a pathological state from other pathological or healthy states. Gene rankings are determined using the integrated gradient method (IG), a neural network-tailored explainability technique, and are compared with 3 commonly-used statistical methods: DESeq2, EdgeR and mutual information (MI). Additionally, an over-representation analysis is conducted to explore the biological relevance of the top-ranked genes. 

% Summary results + interpretation
Several key insights emerge. First, the classifiers consistently achieve a balanced accuracy exceeding 95\% in the majority of cases, with best performance being obtained by LR. Additionally, as cross-validation experiments favoured shallow MLP comprising 1 or 2 layers, this suggests limitations in leveraging complex relationships within the data. 
Second, the classifiers can maintain the same performance using only on a small set of top-ranked genes for training. These genes are potential biomarkers for the studied pathologies. Yet, substantial differences appear in the genes identified across the different methods, highlighting the influence of inherent bias in each gene selection method.

However, good classification performance is also obtained with limited sets of ranked lower, revealing both dispersion and redundancy of information in the gene space. This dispersion is easily measurable for the binary classification tasks with a Welch's unequal variance t-test corrected for multiple testing. This statistical test quantifies the difference between the means of the distributions for each class, accounting for their average variances. Taking the ttg-breast dataset as an example, around 90\% of the genes have an adjusted p-value below 5\%. Most genes carry enough information to distinguish the classes. Hence, in trained models, up to 20~\% of the most important genes ($100-PGI \simeq 20\%$) can be masked without affecting classification performance. Similarly, re-training models on lower-ranked genes achieves good performance. 

To gain insights into the genes preferentially used by ML methods, looking at the correlation of rankings with the t-statistic is informative. 
On the ttg-breast dataset, scores from DESeq2, EdgeR and MI are highly correlated with the t-statistic, with Spearman correlations around $0.9$. LR and MLP exhibit lower correlation around $0.45$. However, on a more complex dataset such as ttg-all, characterised by more heterogeneous data samples, correlations of the ML-based rankings with the t-statistic decrease. Surprisingly, GNN consistently exhibits low correlations, notably for ttg-all where the t-statistic distribution of the top 100 genes selected by IG is similar to those of genes chosen randomly. This highly questions the interpretability of genes by the GNN (IG) method. 

In terms of classification performance, statistical methods looking for differentially expressed genes prove to be even better than ML methods in selecting a minimal set of informative genes (Fig. \ref{fig:classif_perf_MLP}). A hypothesis is that during training, the MLP performance rapidly converges to 100\% of accuracy, making the use of the most most differentially expressed genes unnecessary.

From a biological point of view, the concept of explainability, as defined within the ML community, does not immediately provide a novel perspective to unravel complex biological pathways. Given the dispersed and redundant nature of relevant information, transitioning from individual genes to an exploration of cellular processes holds promise for enhancing our understanding of biological phenomena. Here, the over-representation analysis of the top 100 genes identified by the different methods highlights certain cellular processes, occasionally related to the studied pathologies. While some of them may be shared among methods, there is a substantial variability. The different genes identified by the methods may prove complementary, underscoring distinct processes leading to a phenotype. Further investigation of functional sets of genes, using for instance mechanistic experiments, could help to confirm this hypothesis.

% Main conclusion + next directions
In summary, this study provides valuable insights for researchers seeking biologically relevant molecular signatures of pathologies using ML. The investigation of functional gene sets is a promising research direction.

\section{Competing interests}
No competing interest is declared.

\section{Author contributions statement}
MyB, JA, BA, PB designed the work. MyB, AH, MaB mined the bioinformatics literature. MyB retrieved the datasets and conceived the experiments.
MyB prepared the first draft of the manuscript, the figures and output tables. All authors analysed the result, revised, read and approved the final manuscript.

\section{Acknowledgments}
% GraphNex
Grants: CHIST-ERA-19-XAI-006 and GRAPHNEX ANR-21-CHR4-0009.
% CBP
We thank the Centre Blaise Pascal's IT test platform  (ENS de Lyon, France) for the ML facilities~\cite{quemener2013sidus}.

\bibliographystyle{unsrt}
{\small \bibliography{reference}}
\vfill 

\newpage

\beginsupplement

\onecolumn

\section{Supplementary material}
% Tables 4 to 6 and Figures 7 to 11 are here. 

\begin{table*}[h]
\centering
\begin{minipage}[t]{0.99\textwidth}
\centering
\begin{tabular}{|c|c|c|c|c|c|c|c|}
\hline
\multirow{2}{*}{Datasets} & \multicolumn{2}{c|}{LR} & \multicolumn{2}{c|}{MLP} & \multicolumn{3}{c|}{GNN} \\
\cline{2-8}
 & \small{$\lambda$ (L1)} & \small{$\lambda$ (L2)} & Lay. & Feat. & Lay. & Feat. & $k$\\
\hline
PanCan & 1 & 0.1 & 1 & 20 & 1 & 2 & 2\\
BRCA & 0.1 & 0.01 & 1 & 20 & 1 & 2 & 2\\
\small{BRCA-pam} & 0.1 & 1 & 1 & 20 & 1 & 2 & 10\\
ttg-breast & 0.1 & 0.1 & 1 & 20 & 1 & 1 & 10\\
ttg-all & 1 & 0.1 & 2 & 40 & 1 & 2 & 2\\
\hline
\end{tabular}
\captionof{table}{Hyperparameters selected by grid search: the weight $\lambda$ for the LR regularisation term (for trade-off with the cross-entropy loss) ($\lambda$ chosen as 0.01, 0.1, 1 or 10); the number of MLP layers (1 or 2), their number of hidden features (10, 20, 40 or 80); the number of GNN layers (1, 3 or 5), their number of hidden features (1 or 2), and $k$ setting the number of graph edges to $k \times G$ ($k$ is 2 or 10).
\vspace*{-0.5cm}
}
\label{tab:data_hyperparameter}
\end{minipage}
\end{table*}

\begin{minipage}[t]{0.49\columnwidth}
\centering
\begin{tabular}{|c|c|c|c|c|}
\hline
Dataset & LR+L1 & LR+L2 & MLP & GNN\\
\hline
PanCan & $96.6$ & $96.2$ & $96.0 \pm 0.2$ & $94.5 \pm 0.2$\\
BRCA & $99.4$ & $97.3$ & $99.3 \pm 0.1$ & $99.2 \pm 0.1$\\
BRCA-pam & $91.8$ & $89.5 \pm 0.2$ & $87.7 \pm 1.3$ & $87.5 \pm 1.3$\\
ttg-breast & $99.5$ & $98.7 \pm 0.1$ & $99.3 \pm 0.2$ & $99.2 \pm 0.1$\\
ttg-all & $99.5$ & $99.5$ & $99.6$ & $99.4 \pm 0.1$\\
\hline
\end{tabular}
\captionof{table}{Classification performance measured by accuracy (\%). Standard deviations are computed from 10 replicates; not reported when below 0.05.}
\label{tab:classif_perf2}
\end{minipage}
\hspace{1.5cm}
\begin{minipage}[t]{0.32\textwidth}
\centering
\begin{tabular}{|c|c|c|c|c|}
\hline
Dataset & LR+L1 & LR+L2 & MLP & GNN\\
\hline
PanCan & 58253 & 42564 & 74 & 404\\
BRCA & 222 & 175 & 6 & 8\\
BRCA-pam & 673 & 496 & 7 & 11\\
ttg-breast & 262 & 207 & 10 & 12\\
ttg-all & 6127 & 2703 & 136 & 150\\
\hline
\end{tabular}
\captionof{table}{Average training duration (s) for each model.}
\label{tab:classif_time2}
\end{minipage}

\begin{figure*}[h]
\input{figures/heatmap2}
\caption{Heatmaps showing the percentage of common genes among the top 10 (lower) and top 100 (upper + diagonal) genes selected by each method.}
\label{fig:heatmaps2}
\end{figure*}

\begin{figure*}[h]
\centering
\begin{tikzpicture}
\begin{groupplot}[group style={group size=4 by 1, horizontal sep=0.6cm, vertical sep=1.2cm}]

    \nextgroupplot[
        width=0.29\textwidth, 
        height=0.19\textwidth,
        ylabel style={align=center},
        xlabel style={align=center},
        xlabel={\small{Proportion of genes (\%)}\\(a) BRCA},
        yticklabels={LR+L1, LR+L2, MLP, GNN},
        ytick={1,2,3,4},
        ylabel near ticks,
	  xmax=55,
        xmin=-5,
	    ]

     \addplot[orange, thick, only marks, mark=o, error bars/.cd, x dir=both, x explicit] table [y=Index, x expr=100 -\thisrow{Value}, col sep=comma, x error=Std]{csv/evaluate/BRCA/PGI_global_train.csv};

     \addplot[jublue1, thick, only marks, mark=o, error bars/.cd, x dir=both, x explicit] table [y=Index, x=Value, col sep=comma, x error=Std]{csv/evaluate/BRCA/PGU_global_train.csv};

    \nextgroupplot[
        width=0.29\textwidth, 
        height=0.19\textwidth,
        ylabel style={align=center},
        xlabel style={align=center},
        xlabel={\small{Proportion of genes (\%)}\\(b) ttg-all},
        yticklabels={, , , , },
        ytick={1,2,3,4},
        ylabel near ticks,
	  xmax=85,
        xmin=-5,
	    ]

     \addplot[orange, thick, only marks, mark=o, error bars/.cd, x dir=both, x explicit] table [y=Index, x expr=100 -\thisrow{Value}, col sep=comma, x error=Std]{csv/evaluate/ttg-all/PGI_global_train.csv};

     \addplot[jublue1, thick, only marks, mark=o, error bars/.cd, x dir=both, x explicit] table [y=Index, x=Value, col sep=comma, x error=Std]{csv/evaluate/ttg-all/PGU_global_train.csv};

    \nextgroupplot[
        width=0.29\textwidth, 
        height=0.19\textwidth,
        ylabel style={align=center},
        xlabel style={align=center},
        xlabel={\small{Proportion of genes (\%)}\\(c) PanCan},
        yticklabels={, , , , },
        ytick={1,2,3,4},
        ylabel near ticks,
	  xmax=105,
        xmin=-5,
	    ]

     \addplot[orange, thick, only marks, mark=o, error bars/.cd, x dir=both, x explicit] table [y=Index, x expr=100 -\thisrow{Value}, col sep=comma, x error=Std]{csv/evaluate/pancan/PGI_global_train.csv};

     \addplot[jublue1, thick, only marks, mark=o, error bars/.cd, x dir=both, x explicit] table [y=Index, x=Value, col sep=comma, x error=Std]{csv/evaluate/pancan/PGU_global_train.csv};

    \nextgroupplot[
        width=0.29\textwidth, 
        height=0.19\textwidth,
        hide axis,
        xmin=0,
        xmax=100,
        ymin=0,
        ymax=100,
        legend style={at={(0.2, 0.15)}, anchor=south west, /tikz/every even column/.append style={column sep=1cm}}
        ]

    \addlegendimage{jublue1, thick, mark=o}
    \addlegendentry{PGU} 
    \addlegendimage{orange, thick, mark=o}
    \addlegendentry{100 - PGI}

\end{groupplot}
\end{tikzpicture}
\caption{Impact of progressive gene masking on the predictions of ML models. Genes are masked by increasing (PGU) or decreasing importance (PGI) using $\boldsymbol{\phi}^\text{IG}$ scores. PGs are averaged over all training samples correctly classified without masking, with error bars representing standard deviations.
\vspace*{-1cm}
}
\label{fig:PGs2}
\end{figure*}
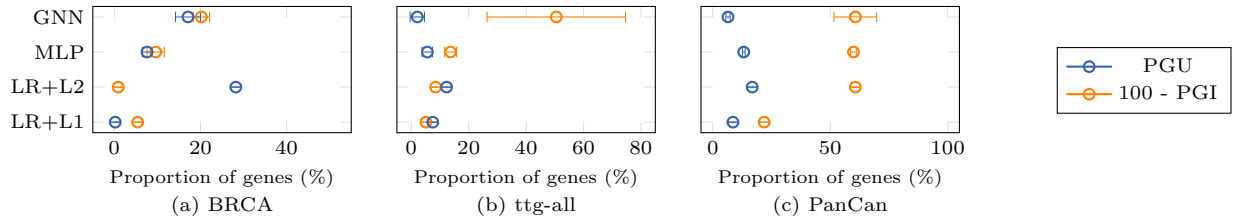

\begin{figure*}[h]
\centering
\input{figures/GSEA2}
\caption{Top 10 over-represented gene sets identified by the different methods (blue dots). When one of these gene sets appear among the top 100 over-represented sets of another method, it is flagged with a light green dot.}
\label{fig:GSEA2}
\end{figure*}

%USE THE BELOW OPTIONS IN CASE YOU NEED AUTHOR YEAR FORMAT.
%\bibliographystyle{abbrvnat}
%\bibliography{reference}

%% sample for biography with author's image
%\begin{biography}{{\color{black!20}\rule{77pt}{77pt}}}{\author{Author Name.} This is sample author biography text. The values provided in the optional argument are meant for sample purposes. There is no need to include the width and height of an image in the optional argument for live articles. This is sample author biography text this is sample author biography text this is sample author biography text this is sample author biography text this is sample author biography text this is sample author biography text this is sample author biography text this is sample author biography text.}
%\end{biography}

%% sample for biography without author's image
%\begin{biography}{}{\author{Author Name.} This is sample author biography text this is sample author biography text this is sample author biography text this is sample author biography text this is sample author biography text this is sample author biography text this is sample author biography text this is sample author biography text.}
%\end{biography}

\end{document}